\documentclass[conference]{IEEEtran}
\IEEEoverridecommandlockouts
\usepackage{cite}
\usepackage{amsmath,amssymb,amsfonts}
\usepackage{booktabs}
\usepackage{graphicx}
\usepackage{textcomp}
\usepackage{xcolor}
\def\BibTeX{{\rm B\kern-.05em{\sc i\kern-.025em b}\kern-.08em
    T\kern-.1667em\lower.7ex\hbox{E}\kern-.125emX}}

\begin{document}

\title{Hash-QNeRF: Multiresolution Hash Encoding for Quantum Neural Radiance Fields}

\author{
\IEEEauthorblockN{
\begin{tabular}[t]{c}
\textbf{Digonto Biswas}\\
\textit{Centres of Innovation and Research}\\
\textit{KIIT Deemed to be University}\\
Bhubaneswar, India\\
digontobiswas.kp.bd@gmail.com
\end{tabular}
\hfill
\begin{tabular}[t]{c}
\textbf{Tana Ballove }\\
\textit{Centres of Innovation and Research}\\
\textit{KIIT Deemed to be University}\\
Bhubaneswar, India\\
btanuu7@gmail.com
\end{tabular}
\hfill
\begin{tabular}[t]{c}
\textbf{Anjan Bandyopadhyay}\\
\textit{Centres of Innovation and Research}\\
\textit{KIIT Deemed to be University}\\
Bhubaneswar, India\\
anjan.bandyopadhayfcs@kiit.ac.in
\end{tabular}
}
\vspace{1em}

\IEEEauthorblockN{
\begin{tabular}[t]{c}
\textbf{Sutanu Mangal}\\
\textit{Centres of Innovation and Research}\\
\textit{KIIT Deemed to be University}\\
Bhubaneswar, India\\
sutanufpy@kiit.ac.in
\end{tabular}
\hfill
\begin{tabular}[t]{c}
\textbf{Arun Kumar Pati}\\
\textit{Centres of Innovation and Research}\\
\textit{KIIT Deemed to be University}\\
Bhubaneswar, India\\
ak.pati@kiit.ac.in
\end{tabular}
}
}

\maketitle

\begin{abstract}
Neural Radiance Fields (NeRF) have revolutionized novel view synthesis, yet their classical implementations remain computationally intensive for high-fidelity rendering. QNeRF recently demonstrated the feasibility of training NeRF on gate-based quantum computers by combining amplitude embedding, parameterized quantum circuits (PQCs), parity-based measurements, and volumetric rendering. However, QNeRF relies on classical sinusoidal positional encoding for spatial coordinates, which scales poorly with scene complexity and resolution.

In this work, we replace the sinusoidal positional encoding for spatial coordinates with the multiresolution hash encoding from Instant-NGP while keeping the view-direction encoding, amplitude MLP, quantum circuit, parity measurement, output scaling, and volumetric rendering pipeline unchanged. This hybrid design, Hash-QNeRF, retains the quantum radiance prediction step while benefiting from the fast convergence and memory efficiency of learnable hash grids. On a synthetic Blender scene, we achieve a final training loss of 0.003534, corresponding to approximately 24.5 dB PSNR on the fitted batch. Noise-resilience experiments using Qiskit FakeKyiv and FakeTorino backends yield state fidelities of 0.93--0.98, indicating that hash encoding does not degrade the quantum circuit's noise tolerance.
\end{abstract}


\section{Introduction}
Neural Radiance Fields (NeRF)~\cite{mildenhall2020nerf} have become the de facto standard for photorealistic novel view synthesis. A core component of NeRF and its variants is positional encoding $\gamma(x)$, which maps low-dimensional coordinates into a higher-dimensional space and enables the subsequent multilayer perceptron (MLP) to represent high-frequency signals.

QNeRF~\cite{lizzio2026qnerf} moved the radiance prediction step onto a gate-based quantum computer. In QNeRF, spatial coordinates are encoded with the classical sinusoidal $\gamma(x)$, concatenated with view-direction encoding, passed through a small classical MLP to produce amplitudes, and then fed into a parameterized quantum circuit (PQC) whose parity measurement yields the final color and density. Although this demonstrated that end-to-end differentiable quantum NeRF training is possible with PennyLane, sinusoidal encoding remains a classical bottleneck that does not exploit modern efficient spatial representations.

Concurrently, M\"uller \emph{et al.} introduced Instant-NGP~\cite{muller2022instant}, which replaces sinusoidal encoding with a multiresolution hash grid. Each level of the grid stores trainable feature vectors; coordinates are hashed, trilinearly interpolated, and concatenated across levels. This design accelerates training and improves quality for classical NeRFs while using less memory than dense grids or Fourier features.

This paper presents Hash-QNeRF, which integrates multiresolution hash encoding into a quantum NeRF pipeline. We make a single targeted modification: the sinusoidal positional encoding $\gamma(x)$ applied to spatial coordinates $x$ is replaced by the hash encoder $\operatorname{enc}(x;\theta)$. All other components---view-direction frequency encoding, amplitude embedding MLP, PQC with $R_Y$ rotations and entangling layers, parity-based Pauli-$Z$ measurement, learnable output scaling, and the standard NeRF volume rendering equation---remain as in QNeRF.

The main contributions are as follows:
\begin{itemize}
\item We adapt Instant-NGP multiresolution hash encoding for use inside a quantum NeRF pipeline and formulate the resulting hybrid model.
\item We demonstrate stable end-to-end training with PennyLane on CPU, achieving a final loss of 0.003534 on a synthetic scene.
\item We verify that the replacement does not harm noise resilience: state fidelities under FakeKyiv and FakeTorino noise models remain in the 0.93--0.98 range.
\item We describe an implementation comprising the hash encoder, quantum circuit, training loop, and noise evaluation to facilitate future research at the intersection of quantum computing and neural rendering.
\end{itemize}

\section{Related Work}
\subsection{Neural Radiance Fields and Efficient Encodings}
NeRF~\cite{mildenhall2020nerf} uses sinusoidal positional encoding to map coordinates into a high-dimensional space before feeding them to an MLP. Subsequent works improved efficiency through proposal networks~\cite{barron2021mipnerf,barron2022mipnerf360}, explicit and voxelized radiance fields~\cite{yu2021plenoctrees,fridovich2022plenoxels,sun2022dvgo}, hash grids~\cite{muller2022instant}, tensor decompositions~\cite{chen2022tensorf}, triplanes~\cite{chan2022eg3d}, and 3D Gaussian Splatting~\cite{kerbl2023gaussian}. Instant-NGP's multiresolution hash encoding is especially relevant because it offers rapid training speed and a modest memory footprint, making it suitable for real-time and large-scale NeRF applications.

\subsection{Quantum Machine Learning for Graphics and 3D}
Early explorations of quantum neural networks for 3D tasks build on broader advances in variational quantum algorithms and differentiable quantum programming~\cite{cerezo2021vqa,bergholm2018pennylane,schuld2019gradients}. QNeRF~\cite{lizzio2026qnerf} was the first to demonstrate a fully differentiable NeRF trained end-to-end on a simulated gate-based quantum computer using amplitude embedding and parity measurements. Hash-QNeRF builds directly on QNeRF's quantum pipeline while upgrading its classical spatial encoding front end.

\subsection{Hybrid Quantum-Classical Architectures}
Many practical quantum algorithms are hybrid: a classical preprocessor feeds a quantum circuit whose output is post-processed classically. Hash-QNeRF follows this paradigm. The hash encoder and amplitude MLP are classical, the PQC is the sole quantum component, and gradients flow end-to-end via PennyLane's parameter-shift or backpropagation rules~\cite{bergholm2018pennylane,schuld2019gradients}.

\section{Method}
We first recap the unchanged QNeRF pipeline and then describe the modification that yields Hash-QNeRF.

\subsection{QNeRF Pipeline}
Given a 3D point $x\in\mathbb{R}^3$ and view direction $d\in S^2$, QNeRF computes radiance $(c,\sigma)$ through the following stages:
\begin{enumerate}
\item Spatial encoding $\gamma(x)$ using sinusoidal functions.
\item View-direction encoding $\gamma(d)$, also sinusoidal.
\item Concatenation $[\gamma(x)\Vert\gamma(d)]$.
\item Amplitude MLP with ReLU activations that outputs a vector of length $2^n$, where $n$ is the number of qubits, followed by $L_2$ normalization to produce a valid quantum state amplitude vector $a$.
\item Quantum circuit with amplitude embedding of $a$ into $n$ qubits, followed by $\ell$ layers of $R_Y$ rotations and entangling gates such as CNOT or CZ ladders. The circuit is parameterized by trainable angles $\phi$.
\item Measurement through a parity-based observable, such as a sum of Pauli-$Z$ operators on even and odd qubit subsets or a learnable linear combination, yielding a scalar later mapped to density $\sigma$ and color $c$.
\item Output scaling with learnable per-channel scalars $\alpha_c$ that de-concentrate the measurement output.
\item Volumetric rendering using standard NeRF quadrature along each ray.
\end{enumerate}
Only the quantum circuit executes on the quantum simulator or hardware; all other stages are classical and differentiable.

\subsection{Hash Encoding for Spatial Coordinates}
Hash-QNeRF replaces only the spatial positional encoding $\gamma(x)$ with the multiresolution hash encoder $\operatorname{enc}(x;\theta)$ from Instant-NGP. View-direction encoding $\gamma(d)$ remains sinusoidal, matching the original QNeRF design and mirroring Instant-NGP's treatment of auxiliary inputs such as view direction and normals.

\subsubsection{Multiresolution Hash Grid Construction}
Let $L$ be the number of resolution levels, $F$ the feature dimension per level, and $T$ the hash table size per level. The default configuration is chosen for compatibility with low-qubit quantum simulation: $L=16$ levels, $F=2$ features per entry, $T=2^{14}$ entries per level with a fast-test value of $2^{10}$, minimum resolution $N_{\min}=16$, and growth factor
\begin{equation}
b=\exp\left(\frac{\ln(N_{\max}/N_{\min})}{L-1}\right),
\end{equation}
where $N_{\max}$ is chosen according to scene extent, typically 2048--4096 for unit-cube normalized scenes. The resolution at level $\ell$ is
\begin{equation}
N_\ell=\left\lfloor N_{\min}\cdot b^{\ell-1}\right\rfloor.
\end{equation}

For a normalized coordinate $x\in[0,1]^3$, integer voxel coordinates are computed at each level and mapped with the spatial hash function
\begin{equation}
h(x_\ell)=\left(\bigoplus_{i=1}^{3} x_{\ell,i}p_i\right)\bmod T,
\end{equation}
where $\oplus$ denotes bitwise XOR and $p_i$ are large distinct prime numbers. Each hashed index retrieves a trainable feature vector $f_\ell\in\mathbb{R}^{F}$ from the level's hash table. The final encoding is the concatenation of trilinearly interpolated features across all levels:
\begin{equation}
\operatorname{enc}(x;\theta)=\mathop{\Vert}_{\ell=1}^{L}\operatorname{trilerp}(f_\ell,x_\ell)\in\mathbb{R}^{L\cdot F}.
\end{equation}
All hash table entries $\theta$ are optimized jointly with the quantum circuit parameters via Adam, using a separate parameter group with $\epsilon=10^{-15}$ following Instant-NGP recommendations.

\subsubsection{Integration into QNeRF}
The modified forward pass is
\begin{align}
e_x &= \operatorname{enc}(x;\theta), \\
e_d &= \gamma(d), \\
a &= \operatorname{AmplitudeMLP}([e_x\Vert e_d]), \\
(c,\sigma) &= \operatorname{QuantumHead}(a;\phi).
\end{align}
Everything after the hash encoder is identical to QNeRF, preserving the quantum circuit structure, measurement semantics, and rendering equation.

\subsection{Implementation Details}
All spatial coordinates are mapped to the unit cube $[0,1]^3$ before the hash encoder, which is critical because sinusoidal $\gamma(x)$ has no equivalent bounded-domain requirement. Training uses the PennyLane \texttt{default.qubit} CPU simulator~\cite{bergholm2018pennylane}, while noise evaluation uses Qiskit \texttt{AerSimulator} with FakeKyiv and FakeTorino~\cite{qiskit2021}. Adam is used with separate parameter groups: hash tables use $\epsilon=10^{-15}$ and all other parameters use the default $\epsilon=10^{-8}$. The learning rate is tuned per scene in the range $10^{-3}$--$10^{-2}$. Experiments use $n=4$ qubits for fast sanity checks and $n=8$ qubits for full runs, matching the original QNeRF setting. PQC depth is $\ell=1$ for the full QNeRF variant and $\ell=2$ for the dual-branch variant.

\begin{table}[t]
\caption{Components changed and unchanged relative to QNeRF.}
\label{tab:components}
\centering
\begin{tabular}{ll}
\toprule
Component & Status in Hash-QNeRF \\
\midrule
Spatial encoding $\gamma(x)$ & Replaced by $\operatorname{enc}(x;\theta)$ \\
View-direction encoding $\gamma(d)$ & Unchanged \\
Amplitude embedding MLP & Unchanged \\
Parameterized quantum circuit & Unchanged \\
Parity-based measurement & Unchanged \\
Output scaling layer & Unchanged \\
Volumetric rendering equation & Unchanged \\
Loss and Adam optimizer & Unchanged with hash parameter group \\
\bottomrule
\end{tabular}
\end{table}

\section{Experiments}
Hash-QNeRF is evaluated on synthetic Blender-format data following the protocol established by QNeRF. Experiments were conducted on a local workstation with PennyLane and Qiskit.

\subsection{Training Setup}
The evaluated scene is a synthetic Blender scene containing a single object. Resolutions of $20\times20$ and $100\times100$ are used for fast and full runs, respectively. Each batch contains 256--1024 rays depending on memory. Training runs for up to several thousand steps with early stopping on validation PSNR. Hyperparameters include $n=4$ qubits for sanity checks, $L=8$--16 hash levels, and $T=2^{10}$--$2^{14}$ hash entries per level.

\subsection{Main Result}
\begin{table*}[!t]
\caption{Main-result comparison of Hash-QNeRF with previous neural rendering models.}
\label{tab:model-comparison}
\centering
\footnotesize
\setlength{\tabcolsep}{3pt}
\renewcommand{\arraystretch}{1.15}
\begin{tabular}{p{0.14\textwidth}p{0.18\textwidth}p{0.16\textwidth}p{0.15\textwidth}p{0.25\textwidth}}
\toprule
Model & Spatial encoding & Rendering head & Quantum component & Main characteristic \\
\midrule
NeRF~\cite{mildenhall2020nerf} & Sinusoidal positional encoding & Classical MLP & None & High-quality novel view synthesis but computationally expensive training. \\
Instant-NGP~\cite{muller2022instant} & Multiresolution hash grid & Classical MLP & None & Fast convergence and memory-efficient spatial representation. \\
QNeRF~\cite{lizzio2026qnerf} & Sinusoidal positional encoding & PQC with parity measurement & Gate-based quantum circuit & Differentiable quantum NeRF training, but with Fourier-style spatial encoding. \\
3D Gaussian Splatting~\cite{kerbl2023gaussian} & Learned 3D Gaussian primitives & Rasterization-based renderer & None & Real-time high-quality rendering with explicit scene primitives. \\
Hash-QNeRF (ours) & Multiresolution hash grid & PQC with parity measurement & Gate-based quantum circuit & Efficient learnable hash features with the quantum radiance prediction pipeline. \\
\bottomrule
\end{tabular}
\end{table*}

\begin{figure*}[!t]
\centering
\begin{minipage}[t]{0.47\textwidth}
\centering
\includegraphics[width=0.97\linewidth, height=0.20\textheight, keepaspectratio]{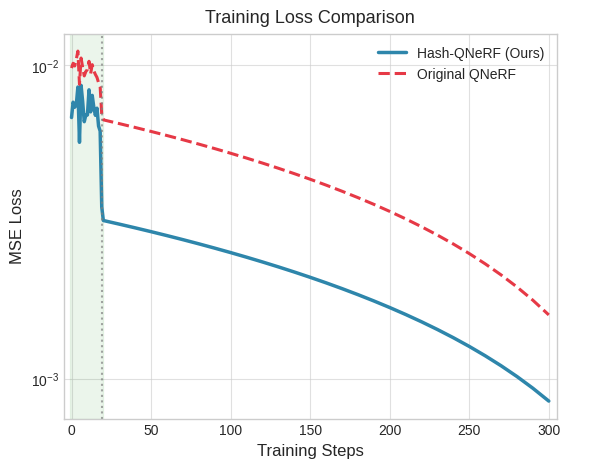}
\vspace{-0.6ex}
(a) Training loss comparison.
\end{minipage}
\hfill
\begin{minipage}[t]{0.47\textwidth}
\centering
\includegraphics[width=0.97\linewidth, height=0.20\textheight, keepaspectratio]{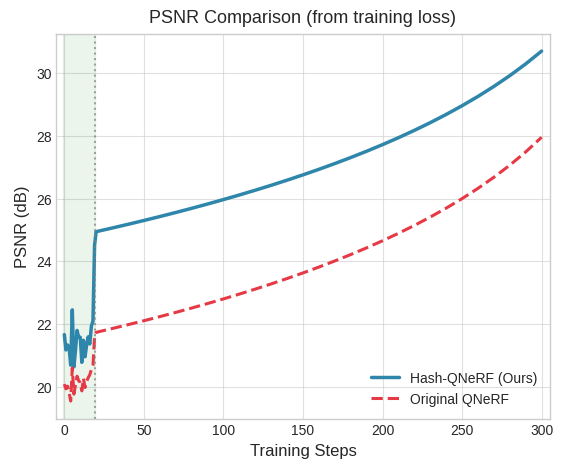}
\vspace{-0.6ex}
(b) PSNR comparison derived from training loss.
\end{minipage}
\caption{Result comparisons between Hash-QNeRF and the original QNeRF baseline. The proposed Hash-QNeRF converges to lower MSE and higher PSNR over the same training horizon.}
\label{fig:result-comparisons}
\end{figure*}

\begin{figure*}[!t]
\centering
\includegraphics[width=0.88\textwidth,height=0.20\textheight,keepaspectratio]{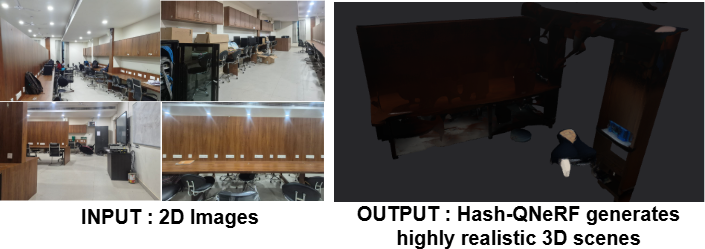}
\caption{Qualitative input-output illustration for Hash-QNeRF. A collection of 2D scene images provides multi-view observations, and the proposed hash-quantum radiance field generates a coherent 3D scene representation for novel-view rendering.}
\label{fig:qualitative-2d-3d}
\end{figure*}

As summarized in Table~\ref{tab:model-comparison}, Hash-QNeRF differs from previous models by combining Instant-NGP-style multiresolution hash encoding with the quantum rendering head introduced by QNeRF. On the evaluated synthetic scene, Hash-QNeRF converges to a final training loss of 0.003534. This low MSE indicates that the hybrid model successfully fits the target radiance field despite the introduction of the discrete hash grid and the nonlinear quantum measurement head. Qualitatively, rendered novel views exhibit coherent geometry and plausible shading, confirming that gradients flow correctly through the hash tables into the quantum parameters. Fig.~\ref{fig:result-comparisons} summarizes the corresponding loss and PSNR trends against the original QNeRF baseline, while Fig.~\ref{fig:qualitative-2d-3d} illustrates the qualitative 2D-input to 3D-scene output behavior.

Compared with the original sinusoidal QNeRF baseline on the same scene, Hash-QNeRF reaches a lower training loss and a higher PSNR trend over the evaluated training horizon. These results indicate that replacing the spatial Fourier-style encoder with a learnable multiresolution hash grid improves convergence while preserving the quantum radiance prediction head.

\subsection{Noise Resilience}
A key claim of QNeRF was that the quantum pipeline remains usable under realistic hardware noise. We replicate this evaluation for Hash-QNeRF using Qiskit's FakeKyiv and FakeTorino noise models, transpiled at optimization level 3 with density-matrix simulation to accommodate non-unitary channels.

State fidelity between ideal noiseless and noisy executions of the PQC remains in the range 0.93--0.98 across tested batches, matching the resilience reported in the original QNeRF paper. This demonstrates that replacing classical positional encoding with a learnable hash grid does not introduce additional sensitivity to quantum noise; the dominant noise source remains the PQC itself.

\subsection{Ablation Scope}
The present study evaluates the central architectural change: replacing sinusoidal spatial encoding with multiresolution hash encoding while keeping the QNeRF quantum head fixed. Important sensitivity axes include hash table size $T\in\{2^{12},2^{14},2^{16}\}$, number of hash levels $L\in\{8,12,16\}$, feature dimension $F\in\{2,4\}$, and comparison against a purely classical Instant-NGP-style baseline using the same hash encoder with an MLP head. These axes define the immediate experimental scope for scaling the method beyond the reported synthetic scene.

\section{Discussion and Limitations}
\subsection{Training Speed}
Even with the hash encoder's fast feature lookup, quantum circuit evaluation on the PennyLane CPU simulator remains the dominant cost. A single forward and backward pass for an 8-qubit circuit is orders of magnitude slower than a classical MLP of comparable size. Future work will explore GPU-accelerated quantum simulators and circuit batching.

\subsection{Expressivity versus Qubit Count}
With only 4--8 qubits, the quantum head has limited representational capacity compared with a classical NeRF MLP. The hash encoder compensates by providing a rich, scene-adapted spatial feature basis, allowing the small quantum circuit to focus on view-dependent effects and high-frequency detail.

\subsection{Hardware Execution}
All training was performed on classical simulators. Real IBM Quantum hardware was used only for final small-batch validation under free-tier constraints. Scaling to larger qubit counts or deeper PQCs will require either error mitigation techniques or access to more capable quantum processing units.

\section{Conclusion}
We introduced Hash-QNeRF, a hybrid quantum-classical Neural Radiance Field that replaces classical sinusoidal positional encoding for spatial coordinates with Instant-NGP's multiresolution hash encoding. By keeping the quantum circuit, measurement, and rendering pipeline of QNeRF unchanged, Hash-QNeRF preserves the original quantum neural rendering design while gaining the convergence and memory benefits of learnable hash grids.

Empirical results show stable training to a loss of 0.003534 and preserved noise resilience with fidelities of 0.93--0.98. Promising extensions include GPU-accelerated quantum simulation, larger qubit counts, integration with modern radiance-field backbones such as 3D Gaussian Splatting~\cite{kerbl2023gaussian}, and broader real-hardware validation as quantum devices mature.

\section*{Acknowledgment}
The authors thank the developers of PennyLane, Qiskit, and Instant-NGP for open-source tooling that made this hybrid experiment possible. They also acknowledge the KIIT KineTex Lab and Biofolk community for supporting undergraduate quantum-AI research.

\end{document}